\documentclass[12pt]{article}
\usepackage{epsfig,amsmath,latexsym,amsfonts}

\begin{document}
\font\cmss=cmss10 \font\cmsss=cmss10 at 7pt
\hfill Bicocca-FT-02-25

\vskip .1in \hfill hep-th/0212181

\hfill

\vspace{20pt}

\begin{center}
{\Large \textbf{3-D INTERACTING CFTs AND GENERALIZED HIGGS PHENOMENON IN 
HIGHER SPIN THEORIES ON AdS}}
\end{center}

\vspace{6pt}

\begin{center}
\textsl{L. Girardello $^{a}$, M. Porrati $^{b,c}$, A. Zaffaroni $^{a}$} 
\vspace{20pt}

\textit{$^a$ Universit\`{a} di Milano-Bicocca and INFN\\ Piazza della 
Scienza 3, I-20126 Milano (Italy)}

\vspace{10pt} 

\textit{$^b$ Department of Physics, New York University\\ 4 Washington
Pl., New York NY 10012 (USA)}

\vspace{10pt}

\textit{$^c$ Scuola Normale Superiore\\ Piazza dei Cavalieri 7,
I-56126 Pisa (Italy)}\vspace{10pt}

\end{center}

\vspace{12pt}

\begin{center}
\textbf{Abstract}
\end{center}

\vspace{4pt} {\small \noindent We study a duality, recently
conjectured by Klebanov and Polyakov, between higher-spin theories on
$AdS_4$ and $O(N)$ vector models in 3-d.  These theories are free in
the UV and interacting in the IR.  At the UV fixed point, the $O(N)$
model has an infinite number of higher-spin conserved currents. In the
IR, these currents are no longer conserved for spin $s>2$.  In this
paper, we show that the dual interpretation of this fact is that all
fields of spin $s>2$ in $AdS_4$ become massive by a Higgs mechanism,
that leaves the spin-2 field massless.
We identify the Higgs field and show how it
relates to the RG flow connecting the two CFTs, which is induced by 
a double trace deformation.}
\vfill
\vskip 5.mm
 \hrule width 5.cm
\vskip 2.mm
{\small
\noindent e-mail: girardello@mib.infn.it, massimo.porrati@nyu.edu, 
alberto.zaffaroni@mib.infn.it}

\newpage
\section{Introduction}

Theories with an infinite number of massless higher-spin gauge fields
(HS) have a long story. Recently, they have been reexamined by several
authors~\cite{list}. One of the reasons for this resurgence of
interest is that these theories are candidates for a semi-classical
treatment of the small tension limit of string theory. The important
observation that higher-spin theories can be consistently formulated
in Anti-de-Sitter space~\cite{fv} also suggests that they are useful
in the context of the AdS/CFT correspondence. Weakly coupled gauged
theories contain an infinite number of almost-conserved currents that
may be described by a dual HS theory in Anti de Sitter space.  While
our understanding of the description of the weak coupling limit of
four dimensional YM in terms of higher-spin theories is still elusive,
some progress has been made for certain three dimensional conformal
field theories~\cite{kp}. The specific example of ref.~\cite{kp} deals
with three-dimensional $O(N)$ vector models. The singlet sector of the
$O(N)$ theories contains an infinite number of conserved currents of
even spin in the large $N$ limit. In~\cite{kp}, it was conjectured
that the singlet sector of this theory is dual to that of the higher
spin theories in $AdS_4$ studied by Vasiliev~\cite{v}, which consists
of a single Regge trajectory of even spin \footnote{Further work on
the subject can be found in \cite{tedeschi}.}.  Two different
three-dimensional conformal field theories were considered
in~\cite{kp}, the free $O(N)$ model and the infrared fixed point that
can be obtained by perturbing the free theory with a relevant double
trace operator.  Following the general description of RG flows induced
by double-trace operators~\cite{silverstein,wdouble,berkooz}, the
authors of~\cite{kp} also conjectured that both theories are described
by the same Vasiliev Lagrangian, but with a difference consisting in
the choice of boundary condition for a certain field.  As shown
in~\cite{wdouble}, and further studied in \cite{gubser,gk}, the dual
description of the RG flow induced by double-trace deformation of the
boundary CFT is unusual. Instead of changing the 4-d background, this
flow leaves the geometry unchanged, at least at tree level in the bulk
theory, but it changes the boundary behavior of a certain bulk field.

This raises the question that we want to address in this paper.  When
the bulk Lagrangian is dual to the $O(N)$ model at the IR fixed point,
the higher spin currents are conserved only in the large $N$ limit.
Because of the standard relation between conformal dimension in 3-d,
and mass in $AdS_4$, an infinite number of higher spins should become
massive when $1/N$ corrections are included.  All spins should instead
remain massless in the description of the free UV theory. This raises
an interesting puzzle.  As we said, the two CFTs are described by the
same bulk Lagrangian and they only differ by a choice of boundary
condition of a certain field.  How can such a change of boundary
conditions in a (scalar) field induce masses for all particles of spin
higher than 2?  The answer to this question turns out to be
surprisingly similar to a case recently studied by one of the
authors~\cite{porrati}; namely, a graviton coupled to conformal matter
in $AdS_4$. There, one can show that, when matter is given
non-standard boundary conditions, it can form a {\em bound state} that
acts as the Goldstone vector for the spin 2 field.  In other words, in
that case the graviton gets a mass through a one-loop effect.  In this
note, we show that a similar mechanism can give mass to all
higher-spin fields in the dual of the $O(N)$ model at the IR fixed
point.  The mechanism is intrinsically one-loop in the bulk
theory. That explains naturally why the masses of the higher-spin
fields are $O(1/N)$.  Differently from the case studied in
ref.~\cite{porrati}, here the boundary conditions of the bulk fields
leave the spin-2 field massless, at the fixed points of the the
double-trace RG flow.  We finally show that mass generation can only
occur, for spin $s>2$, when the $AdS_4$ theory is dual to the $O(N)$
model at the IR fixed point.

\section{The AdS/HS Correspondence}
To be self-contained, 
in this section we  briefly review the details of the correspondence 
conjectured in \cite{kp}. 
The $O(N)$ model is formulated in terms of a three-dimensional
scalar transforming in the vectorial representation of $O(N)$, with
Lagrangian:
\begin{equation}
  {\cal L}=\int d^3x \left [\partial \phi^a \partial \phi^a +
{\lambda\over 2N}(\phi^a\phi^a)^2\right],
\label{lag}
\end{equation}
where $a=1,...,N$. The theory has two fixed points. There is an
ultraviolet, free fixed point at $\lambda=0$, and an interacting
infrared fixed point \cite{wilson}. The free UV theory has an infinite
number of conserved currents.  Restricting to operators that are
singlets of $O(N)$ and single trace~\footnote{We follow the misuse of
the term single trace introduced in~\cite{kp}.}, we find a conserved
current for each even spin. We can schematically write it as
\begin{equation}
J_{\mu_1,...,\mu_s}=\phi^a (\overleftrightarrow{\partial})^s\phi^a- 
\mbox{traces}.
\label{curr}
\end{equation}
The IR theory is instead interacting and the currents in
Eq.~(\ref{curr}) are not conserved. We can reasonably assume that the
only conserved current in the IR CFT is the stress-energy
tensor. Among the non-singlet operator there are also other conserved
currents; the Noether currents of the (global) $O(N)$ symmetry.
However, it is known \cite{wilson} that the currents in
Eq.~(\ref{curr}), for $s\ge 2$, also have canonical dimension in the
large $N$ limit
\footnote{For more recent references on the subject see \cite{on}}.
Therefore, for $N=\infty$, the IR theory too has an infinite number of
conserved currents. The currents do acquire anomalous dimensions at
order $1/N$. The conjecture formulated in \cite{kp} states that the
singlet sector of both theories has, in the large $N$ limit, a dual
description in terms of a minimal bosonic HS theory containing one
massless gauge field for each even spin~\cite{v}. In this
correspondence, $N$ must be identified with the inverse cosmological
constant of the HS theory.

The description of the two fixed points differs only by a choice of
boundary conditions.  Let $\Sigma$ be the bulk scalar dual to
$(\phi^a\phi^a)$.  $\phi^a\phi^a$ has dimension 1 in the UV and
dimension $2+O(1/N)$ in the IR \cite{wilson}.  $\Sigma$ is a scalar
field with mass $m^2=-2$ at large $N$ (in units of the cosmological
constant). $\Sigma$ is, therefore, a {\it conformally} coupled scalar
field. The UV and IR conformal dimensions of the operator correspond,
respectively, to the two roots of the equation
\begin{equation}
m^2=\Delta (\Delta-3).
\end{equation}
The quantization of $\Sigma$ is subtle, because both roots
$\Delta_{\pm}$ of the previous equation satisfies the unitary bound
in three dimensions.  The analysis of such cases has been performed
in \cite{kw}. The bulk theory corresponding to the assignment
$\Delta=\Delta_+$ differs from the other, $\Delta=\Delta_-$, by 
boundary conditions. Namely, if $\Sigma$ has asymptotic
behavior
\begin{equation}
\Sigma\sim \alpha z^{\Delta_-} + \beta z^{\Delta_+}, \qquad 
\Delta_+>\Delta_-,
\end{equation}
where $z$ is the AdS radial coordinate, the two possible quantizations are
obtained by interchanging the role of $\alpha$ and $\beta$~\cite{kw}.
In the UV, $(\phi^a\phi^a)$ has dimension $\Delta_-=1$ and it is quantized
with boundary condition $\alpha=0$. In the IR $(\phi^a\phi^a)$ has
dimension $\Delta_+=2$ in the large $N$ limit, and it is quantized with
boundary condition $\beta=0$.

In~\cite{kp} the interesting observation was made that the two CFTs
are connected by a RG flow induced by the double-trace operator
$(\phi^a\phi^a)^2$, which, in the UV, is a dimension-two relevant
operator. An analysis of flows induced by double-trace operators was
carried out in~\cite{silverstein,wdouble,berkooz}. The deformation by
a double-trace operator only modifies the boundary condition for the
bulk field $\Sigma$ dual to $(\phi^a\phi^a)$.  According to
\cite{wdouble}, the boundary condition on $\Sigma$ to be imposed along
the flow is:
\begin{equation}
\alpha=\lambda \beta .
\label{flow}
\end{equation}
We see that in the two limiting cases, $\lambda=0$ and
$\lambda=\infty$, we recover the two different boundary conditions
describing, respectively, the UV and the IR fixed points.

\section{A Generalized Higgs Effect}

It is intriguing to notice that the same Lagrangian gives a
semi-classical description of two different fixed points, one of which
is free while the other is interacting. In particular, an obvious
question can be raised.  Since the currents in Eq.~(\ref{curr}) are
not conserved in the IR at finite $N$, we expect that the
corresponding higher spin fields in the bulk acquire a mass of order
$1/N$, when quantum (loop) corrections are included.  We want now to
prove that this is indeed the case. Namely, that in the bulk
Lagrangian describing the IR fixed point, a generalized Higgs effect
may take place, which gives mass to all fields of spin greater than
two.  We will also show that no Higgs effect is expect in the
Lagrangian describing the UV fixed point, and that boundary conditions
alone are responsible for the different behaviors of the two theories.

In AdS, a spin $s$ field can acquire a mass by ``eating'' a single
massive field of spin $s-1$, by a Higgs-like mechanism.  To describe
this phenomenon properly, recall that the representations of the 3-d
conformal group, $SO(3,2)$, which is also the isometry group of
AdS$_4$, are labeled by their quantum numbers under the maximal
compact subgroup $SO(3)\times U(1)$: the spin $s$ and the conformal
weight $\Delta$. A representation $D(\Delta,s)$ satisfies a shortening
condition when $\Delta =s+1$, and corresponds to a conserved current
in the CFT and a massless field in $AdS_4$.  A massive spin $s$
representation of the conformal group decomposes in the massless limit
as~\cite{he}
\begin{equation}
 D(\Delta,s)\xrightarrow{\Delta\rightarrow s+1} D(s+1,s)\oplus
 D(s+2,s-1) .
\end{equation}
The representation $D(s+2,s-1)$ is the Goldstone field.
Since in $AdS_4$ the energy
spectrum is discrete, two gauge fields can form a bound state with
the quantum number of the Goldstone field (even when they are free!). 
In the presence of an appropriate trilinear coupling,
a spin $s$ field can then acquire mass through radiative corrections.
Let us stress that this phenomenon cannot occur in flat space where
the spectrum is continuous.
This analysis was already performed in~\cite{porrati} in the case of a
graviton in $AdS_4$, coupled to a conformal scalar. 

Let us denote with $W_s\equiv W_{\mu_1,...,\mu_s}$ the spin $s$ gauge
field.  In a Lagrangian as those proposed by Vasiliev~\cite{v}, we
expect many trilinear couplings between gauge fields of different
spin. Some involve the field $\Sigma$ and some do not. Those without
$\Sigma$ can be schematically written as $W^s \partial^k
(W_{s_1}W_{s_2})$ with $s_1+s_2+k=s$, with derivatives arbitrarily
distributed among the gauge fields. They cannot be responsible for the
Higgs mechanism. Even if the product of representations of spin $s_1$
and $s_2$ contains a mode in the representation $D(s+2,s-1)$, the
latter would have wrong parity for being a Goldstone field; let us see
why.  Since we are interested in a one loop effect, we can neglect
$1/N$ corrections to the dimensions of our fields.  All the $W_s$ are
thus massless in the large $N$ limit and, therefore, have dimension
$s+1$.  The ground state of the would-be Goldstone representation has
conformal dimension $s+2$. It is obtained from the lowest weight state
of $D(s_i+1,s_i)$, which has dimension $s_i+1$, by acting on it with
$k$ raising operators of the group $SO(3,2)$.  Since the parity of a
genuine spin $s_i$ field is $(-1)^{s_i}$ and the parity of a raising
operator is $-1$, this mode has parity $P=(-1)^{s_1+s_2+k}=(-1)^s$,
which is the wrong one for a spin $(s-1)$ gauge field.

The field $\Sigma$ is what describes the RG flow, and it is, moreover,
the only one to change under it, to leading order in $1/N$. So, we
expect it to appear in the couplings needed to give mass to our
high-spin fields.  Recall that $\Sigma$ only has different boundary
conditions at the two fixed points of the RG flow.  We can easily
write a trilinear coupling of the form
\begin{equation}
W^{\mu_1,...,\mu_s}W_{\mu_1,...,\mu_{s-2}}\partial_{\mu_{s-1}}\partial_{\mu_s}
\Sigma,
\label{trilinear}
\end{equation}
where, for simplicity, we chose a specific distributions for the
derivatives.  Such coupling can be certainly reconstructed from the
three point function of free fields in the CFT. It is also reminiscent
of the equation for the conservation of the currents in
Eq.~(\ref{curr}). To see this, write the Lagrangian Eq.~(\ref{lag}) in
terms of an auxiliary field $\sigma$,
\begin{equation}
{\cal L}=\int d^3x \left[\partial \phi^a \partial \phi^a +\sigma (\phi^a
\phi^a) - {N \sigma^2\over 2\lambda} \right],
\label{lag2}
\end{equation}
and use the equations of motions $\sigma=\lambda (\phi^a\phi^a)/N$,
$\Box \phi^a=\sigma\phi^a$. Then, the divergence of the current can be 
rewritten, schematically, as 
\begin{equation}
\partial^\mu J_{\mu,\mu_1,...,\mu_{s-1}}\sim 
J_{\mu_1,...,\mu_{s-2}}\partial_{\mu_{s-1}}\sigma \, |_{ST},
\label{noncons}
\end{equation}
where the subscript means that the right hand side is projected on
the symmetric-traceless part. 

The coupling in Eq.~(\ref{trilinear}) can give mass to the 
spin $s$ fields, by a one-loop diagram, only when the product of the 
representations to which $W_{\mu_1,...,\mu_{s-2}}$ and $\Sigma$ belong
contains the Goldstone representation $D(s+2,s-1)$. 
To leading order in $1/N$, 
$\Sigma$ has dimension $\Delta=1$ in the UV, but dimension 
$\Delta=2$ in the IR, while all the $W_s$ have always dimension $s+1$. 
We also have \cite{he}
\begin{equation}
D(s-1,s-2)\oplus D(\Delta,0)=\sum_{S=0}^\infty\sum_{n=0}^\infty 
D(\Delta+S+s+n-1,s+S-2).
\label{rep}
\end{equation}
This equation shows that a mode $D(s+2,s-1)$, with the right quantum
numbers to be the Goldstone, appears for both values of
$\Delta$. However, it is easy to check that the candidate Goldstone
has the same parity of the would-be massive field $W_s$ only when
$\Delta=2$ \footnote{In the case $\Delta=1$, we create the lowest
weights of $D(s+2,s-1)$ by applying two raising operators, $L^{+i}$,
to the product of the lowest weight of $D(s-1,s-2)$ and $D(1,0)$, thus
obtaining a field of parity $P=(-1)^s$. $D(s+2,s-1)$ is thus a {\it
pseudo}-spin $(s-1)$ field.}.  We conclude that, only when $\Sigma$ is
quantized with conformal weight $2$ in the large $N$ limit, a Higgs
mechanism is possible.

We must also check that the graviton remains massless: in a CFT, a
singlet conserved current corresponding to the stress-energy tensor
always exists.  It was already noticed in~\cite{porrati} that the a
graviton coupled to a conformal scalar can acquire mass only if the
boundary conditions on the scalar make it belong to the reducible
representation $D(1,0)\oplus D(2,0)$.  In our case, the scalar belongs
to the $D(1,0)$ in the UV, and to the $D(2,0)$ in the IR, so that no
Higgs mechanism is expected.  We can see this explicitly from the
decomposition
\begin{equation}
D(\Delta^\prime,0)\oplus D(\Delta,0)=\sum_{S=0}^\infty\sum_{n=0}^\infty 
D(\Delta+\Delta^\prime+S+2n,S),
\label{rep2}
\end{equation}
that replaces Eq.~(\ref{rep}) in the case $s=2$. No Goldstone
representation $D(4,1)$ is contained in this formula for
$\Delta=\Delta^\prime$, and $\Delta,\Delta'$ equal to either 1 or 2.

\section{Conclusions}

The duality between $O(N)$ critical vector models and HS theories {\em
\'a la} Vasiliev is still in its infancy. A challenge in establishing
it firmly is that while the 3-d CFT is considerably simpler than in
the adjoint case, the 4-d AdS dual is much more complicated than
semiclassical supergravity.  In this paper we furthered the study of
that duality by showing how to explain a puzzling feature of the IR
(interacting) fixed point of the $O(N)$ model.  There, almost all
higher-spin currents that were conserved in the UV acquire anomalous
dimensions. In the AdS dual, this means that almost all massless
fields of the HS theory become massive.  To interpret this effect as a
Higgs phenomenon, one has to explain how to reconcile it with the fact
that the (double trace) perturbation of the UV theory flowing into the
IR fixed point does not change the AdS background, to leading order in
$1/N$. In this paper, we showed that a radiative Higgs effect, where
the Goldstone particle is composite, can solve this puzzle. We
performed a group theoretical analysis showing that only particles
with spin $s>2$ can become massive, and only at the IR fixed point.
It would be interesting and important to explicitly compute the
one-loop self-energy diagram for all particles in the dual HS
theory~\cite{v}, to check this phenomenon explicitly and
quantitatively.  It may also be possible to extend our analysis to a
model that contains some (or all) the non-singlet currents of $O(N)$,
or to other examples of RG flows in 3d, like those discussed in
\cite{anselmi}.

\subsection*{Acknowledgements}
We would like to thank D. Anselmi for fruitful conversations.
A.Z. would like to thank the Scuola Normale Superiore in Pisa 
for its kind hospitality while this work was done.
L.G. and A.Z. are partially
supported by INFN and MURST under contract 2001-025492, and by 
 the European Commission TMR
program HPRN-CT-2000-00131.
M.P. is supported in part by NSF grant PHY-0070787.

\end{document}